\documentclass[twocolumn]{aastex63}
\renewcommand{\sun}{$_\odot$}

\begin{document}

\title{The JAGWAR Prowls LIGO/Virgo O3 Paper I: 
\\Radio Search of a Possible Multi-Messenger Counterpart of the Binary Black Hole Merger Candidate S191216ap}

\correspondingauthor{D. Bhakta}
\email{deven.r.bhakta@ttu.edu}

\author[0000-0002-7965-3076]{D. Bhakta}
\affil{Department of Physics and Astronomy, Texas Tech University, Box 41051, Lubbock, TX 79409-1051, USA}

\author{K. P. Mooley}
\affil{National Radio Astronomy Observatory, P.O. Box O, Socorro, NM 87801, USA}
\affil{Caltech, 1200 E. California Blvd. MC 249-17, Pasadena, CA 91125, USA}

\author{A. Corsi}
\affil{Department of Physics and Astronomy, Texas Tech University, Box 41051, Lubbock, TX 79409-1051, USA}

\author[0000-0003-0477-7645]{A. Balasubramanian}
\affiliation{Department of Physics and Astronomy, Texas Tech University, Box 41051, Lubbock, TX 79409-1051, USA}

\author[0000-0003-0699-7019]{D.\ Dobie}
\affil{Sydney Institute for Astronomy, School of Physics, University of Sydney, Sydney, New South Wales 2006, Australia.}
\affil{ATNF, CSIRO Astronomy and Space Science, PO Box 76, Epping, New South Wales 1710, Australia}
\affil{ARC Centre of Excellence for Gravitational Wave Discovery (OzGrav), Hawthorn, Victoria, Australia}

\author{D. A. Frail}
\affil{National Radio Astronomy Observatory, P.O. Box O, Socorro, NM 87801, USA}

\author{G. Hallinan}
\affil{Caltech, 1200 E. California Blvd. MC 249-17, Pasadena, CA 91125, USA}

\author[0000-0001-6295-2881]{D. L.\ Kaplan}
\affiliation{Department of Physics, University of Wisconsin-Milwaukee, Milwaukee, WI, USA}

\author{S. T. Myers}
\affil{National Radio Astronomy Observatory, P.O. Box O, Socorro, NM 87801, USA}

\author{L. P. Singer}
\affil{NASA/Goddard Space Flight Center, Greenbelt, MD 20771, USA}


\begin{abstract}
We present a sensitive search with the Karl G. Jansky Very Large Array (VLA) for the radio counterpart of the gravitational wave candidate S191216ap, classified as a binary black hole merger, and suggested to be a possible multi-messenger event, based on the detection of a high energy neutrino and a TeV photon. We carried out a blind search at C band (4--8 GHz) over 0.3 deg$^2$ of the gamma-ray counterpart of S191216ap reported by the High-Altitude Water Cherenkov Observatory (HAWC). Our search, spanning three epochs over 130 days post-merger and having mean source-detection threshold of 75$\mu$Jy/beam (4$\sigma$), yielded 5 variable sources associated with AGN activity and no definitive counterpart of S191216ap.
We find $<$2\% ($3.0\pm1.3$\%) of the persistent radio sources at 6 GHz to be variable on a timescale of $<$1 week (week--months), consistent with previous radio variability studies. 
Our 4$\sigma$ radio luminosity upper limit of $\sim$1.2$\times{10}^{28}$ erg s$^{-1}$ Hz$^{-1}$ on the afterglow of S191216ap, within the HAWC error region, is 5--10 times deeper than previous BBH radio afterglow searches. 
Comparing this upper limit with theoretical expectations given by \citeauthor{Perna2019} for putative jets launched by BBH mergers, for on-axis jets having energy $\simeq10^{49}$ erg, we can rule out jet opening angles $\lesssim$20 degrees (assuming that the counterpart lies within the 1$\sigma$ HAWC region that we observed).


\end{abstract}


%
%
%
%

\section{Introduction}\label{sec:intro}

Recent detections of high energy neutrinos and gravitational waves (GWs) at extragalactic distances \citep{2016PhRvL.116f1102A, 2018Sci...361..147I} have ushered in a new age of ``multi-messenger'' astronomy \citep{2013RvMP...85.1401A,2019BAAS...51c.250B}. The conventional electromagnetic (EM) branch of astronomy has played an important supporting role, helping to pin-point the sources of the neutrinos and GW emission, and to constrain the physical properties of the progenitors \citep[e.g.][]{2017ApJ...848L..12A, 2018ApJ...864...84K}. The study of these multi-messenger events at radio wavelengths has been particularly rewarding. Important highlights include the detection of the first radio afterglow and direct imaging of relativistic outflow from the merger remnant GW170817 \citep{2017Sci...358.1579H,2018Natur.561..355M,2019Sci...363..968G}, and the imaging of the parsec-scale jet of the possible neutrino source TXS\,0506+056 \citep{2019A&A...630A.103B,2020arXiv200500300L}. In this paper we discuss the radio follow-up of S191216ap, the first astrophysical source that may be both a source of GWs and neutrinos. 

S191216ap was first reported as a compact binary merger candidate by the LIGO Scientific Collaboration (LSC) and Virgo Collaboration on 16 December 2019 at 21:33:38.473 UTC \citep{2019GCN.26454....1L}. The GW event was initially classified as a likely ``mass gap'' signal, with one component of the binary having a mass between a definitive neutron star and black hole classification. The event was later re-classified with 99\% probability as a binary black hole \citep[BBH:][]{2019GCN.26570....1L}. 
The final (revised) sky map and distance was posted by the LSC and Virgo, with a 90\% localization region of area 253 deg$^2$ and luminosity distance estimate of 376$\pm$70 Mpc \citep{2019GCN.26505....1L}.

The IceCube Neutrino Observatory reported a single muon neutrino in the direction of S191216ap, 43\,s prior to the GW merger and with an overall p-value of 0.6\% (2.5$\sigma$) and an error radius of only $\pm{4}^\circ$ \citep{2019GCN.26460....1I, 2019GCN.26463....1C}. Initially the High-Altitude Water Cherenkov Observatory (HAWC) reported no candidate gamma-ray events at TeV energies \citep{2019GCN.26455....1H}, but in a re-analysis of their data centered on the IceCube error region the HAWC collaboration reported a sub-threshold event 80\,s after the binary coalescence \citep{2019GCN.26472....1H}. 
This candidate gamma-ray event was found in a 10 sec search and the significance level for this event is 4.6$\sigma$. This corresponds to a gamma-ray flux  
of about $7.3\times10^{-9}$ TeV$^{-1}$~cm$^{-2}$~s$^{-1}$ at 1~TeV for an intrinsic spectrum with an index of $-2$ (Israel Martinez, priv. comm.). 
The coordinates\footnote{We note that these coordinates are outside the 90\% credible region for S191216ap \citep[but within the 98\% credible region;][]{2019GCN.26505....1L}. The mean distance of S191216ap at the HAWC location is $286\pm43$ Mpc.} of the HAWC event are RA: 323.53 deg, Dec: 5.23 deg, with the 68\% containment region (radius) of 0.3 degrees \citep[i.e. 0.28 deg$^2$ region;][]{2019GCN.26472....1H}.


Many high energy observatories were in operation during the S191216ap event, including the ANTARES neutrino detector \citep{2019GCN.26458....1A}, {\it Fermi} GBM \citep{2019GCN.26461....1W}, MAXI GSC \citep{2019GCN.26462....1N}, \textit{Swift}/BAT \citep{2019GCN.26466....1P}, CALET Gamma-Ray Burst Monitor \citep{2019GCN.26481....1S}, AGILE \citep{2019GCN.26486....1V}, AstroSAT \citep{2019GCN.26511....1S}, Insight-HXMT \citep{2019GCN.26569....1L} and Konus-Wind \citep{2020GCN.26835....1R}. A search for events both spatially and temporally coincident with S191216ap failed to find any likely counterpart high energy candidates. A direct quantitative comparison between these non-detections and the HAWC detection is difficult, as most missions did not survey the entire error region, and carried out photon searches in non-overlapping time windows.

\begin{figure*}
\centering
\includegraphics[width=\textwidth]{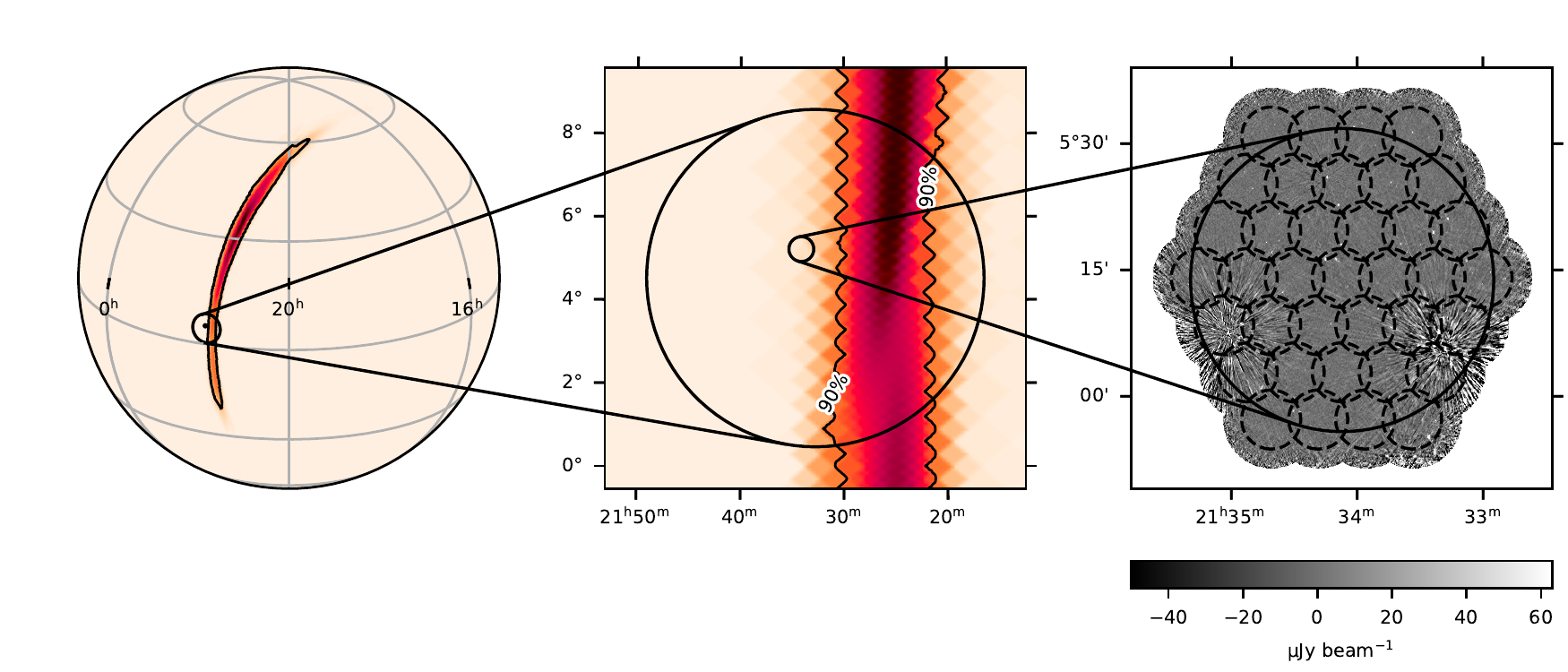}\\
\caption{The LIGO/Virgo sky localization of S191216ap (left) shown along with the localization regions for the IceCube (larger circle, 90\% containment) and HAWC events (smaller circle 0.28 deg$^2$, 68\% containment; middle panel). The right panel shows the sky coverage of our VLA follow up observations of the HAWC 68\% containment region. The 0.38 deg$^2$ image mosaic is from our epoch E3 (see Table~\ref{tab:observations}). The larger black circle corresponds to the 0.28 deg$^2$ HAWC region, while the smaller circles correspond to the primary beams (half-power beam width of about 7 arcminutes) of the 37 pointings at C band (central frequency of 6 GHz). The grey scale bar displays the pixel values, going from $-50~\mu$Jy/beam (black) to 63~$\mu$Jy/beam (white).}
\label{fig:mosaic}
\end{figure*}

This possible multi-messenger event set off a wave of deep searches at X-ray, optical/NIR and radio wavelengths within the first week. Search strategies were of two basic types \citep{2013ApJ...767..124N}, i.e. wide-area and galaxy-targeted searches. Nine galaxies were initially identified in the overlapping LIGO/Virgo-HAWC error region \citep{2019GCN.26479....1S} and within the redshift range of S191216ap, a number that dropped to only three galaxies \citep{2019GCN.26507....1A} after the revised LIGO/Virgo skymap was released \citep{2019GCN.26505....1L}. Targeted searches of these galaxies were made from radio to X-rays wavelengths \citep{2019GCN.26478....1X,2019GCN.26487....1S,2019GCN.26488....1Z,2019GCN.26496....1Y,2019GCN.26530....1M}. Likewise, mosaicked observations were made of all or most of the LIGO-Virgo error region \citep{2019GCN.26464....1A,2019GCN.26473....1L,2019GCN.26488....1Z,2019GCN.26528....1D,2019GCN.26563....1M}, or within the overlap region of LIGO-Virgo and IceCube or HAWC \citep{2019GCN.26475....1E,2019GCN.26483....1R,2019GCN.26498....1E,2019GCN.26509....1O,2019GCN.26531....1M,2019GCN.26605....1S}. With the exception of two optical transients from UKIRT for which no follow-up was undertaken \citep{2019GCN.26605....1S}, there were no compelling EM counterparts identified in that first week. 

Accepting this preliminary identification of S191216ap as a multi-messenger event, we conducted a search for radio counterparts as part of the Jansky VLA mapping of Gravitational Wave bursts as Afterglows in Radio (JAGWAR) program.
In \S\ref{sec:obs_proc} we describe the VLA observations, data processing imaging and source catalog generation.
In \S\ref{sec:var_tran} we describe the search for variable/transient sources, finding no definitive radio counterpart for S191216ap.
We end with a discussion of our results and future prospects for detecting the EM counterparts of BBH mergers.

\section{Observations and Data Processing}\label{sec:obs_proc}


\begin{table*}[htp]
\centering
\scriptsize
\caption{Observing Log}
\label{tab:observations}
\begin{tabular}{llcrccrrr}
\hline\hline 
No. & Start Date  & Epoch & $\Delta$t & Array  & RMS  &  BMAJ & BMIN & BPA \\
    & (UT)        &         & (days) & Config.& ($\mu$Jy/bm)  & (\arcsec) & (\arcsec) & (deg.)  \\
\hline
  1 & 2019-Dec-20 22:47:35 & E1 & 4 & D	    & \\
  2 & 2019-Dec-21 21:49:45 & E1 & 5 & D	    & 18 & 12.3 & 9.4 & 19 \\
  3 & 2019-Dec-22 00:00:36 & E1 & 5 & D	    & \\
  4 & 2019-Dec-27 21:39:31 & E2 & 11 & D	& \\
  5 & 2019-Dec-27 23:50:21 & E2 & 11 & D	& 19 & 11.3 & 9.5 & 16 \\
  6 & 2019-Dec-28 21:22:14 & E2 & 12 & D	& \\
  7 & 2020-Apr-23 13:42:13 & E3 & 129 & C	& \\
  8 & 2020-Apr-23 15:53:03 & E3 & 129 & C	& 14 & 3.9 & 2.9 & 7.5  \\
  9 & 2020-Apr-24 13:38:17 & E3 & 130 & C	& \\
 10 & 2020-Aug-14 07:44:27 & Follow-up & 242 & B & 9--15 & 1.3 & 0.9 & 35 \\
\hline
\multicolumn{9}{p{4.5in}}{Notes: The start date provides the time and date when each observation took place, with $\Delta$t reporting the number of days from the merger. We also report the RMS ($\mu$Jy/bm) for each epoch and the dimensions of the synthesized beam.} 
\end{tabular}
\end{table*}

\subsection{VLA Observations}
With consideration of the HAWC sub-threshold event, we chose to conduct deep C band (4--8 GHz) observations of the gamma-ray 68\% containment region. 
To maximize the continuum imaging sensitivity, we used the Wide-band Interferometric Digital Architecture (WIDAR) correlator with 32 spectral windows, 64 2-MHz-wide channels each to get 4 GHz of total bandwidth centered on 6.0 GHz. 
Our observations were carried out across 3 epochs (E1, E2, E3), with each epoch being divided into 3 observations (for $\sqrt{3}$ improvement in sensitivity), with the Karl G. Jansky Very Large Array (VLA) in C and D array configurations, under the JAGWAR large program (VLA/18B-320; PI: Frail). The epoch time frame ranged from 5 days post merger to 4 months post merger (subject to scheduling constraints and sampling the putative afterglow light curve in logarithmic time-steps). Each observation lasted for 3.6 hr and consisted of 37 pointings, with the goal of creating a standard pointed image mosaic of 0.38 deg$^2$, and achieve fairly uniform sensitivity across the 0.28 deg$^2$ HAWC region. The mosaic is centered on the coordinates reported for the HAWC sub-threshold event \citep{2019GCN.26472....1H}. 3C48 was used as the flux density and bandpass calibrator. The phase calibrator J2130+0502 was observed for a duration of 1 minute every 20--30 minutes.  The observational parameters for all three epochs are listed in Table~\ref{tab:observations} for which we list for each epoch, the array configuration, the rms noise (RMS) and the synthesized beam (BMAJ,BMIN, BPA) of each final image. Figure~\ref{fig:mosaic} shows the image mosaic along with the locations of the VLA pointings and the HAWC 68\% confidence region.

We conducted follow-up observations in C-band of any significant variable sources identified with the mosaicked region. These pointed observations were carried out on 2020 August 14 with the VLA in B array configuration (see Table~\ref{tab:observations}). Integration times varied from $\approx$4-10 minutes, depending on the flux density of the sources. The total duration of the observation was 47 minutes. 3C48 was used as the flux density and bandpass calibrator, while the phase calibrator was J2130+0502. 


\subsection{RFI flagging, Calibration and Imaging}\label{sec:obs_proc.cal}
Directly after the observations for each epoch were completed, we downloaded the raw data from the VLA archive onto the \textit{Lustre} file system at the NRAO AOC in Socorro. The raw data was then calibrated using the NRAO CASA pipeline (in CASA 5.6.1). 
Post-calibration, we carried out manual flagging to remove spectral windows affected by residual RFI. For the imaging, we used CASA {\tt clean} with Briggs weighting (robust factor of 0.5), two Taylor terms, and a threshold of 0.1 mJy. The pixel size was chosen to sample the synthesized beam across 6 pixels for the first two epochs and 4 pixels for the third epoch. 
The central frequency for each image is 6.0 GHz. 

\begin{figure}
\centering
\includegraphics[width=3.5in]{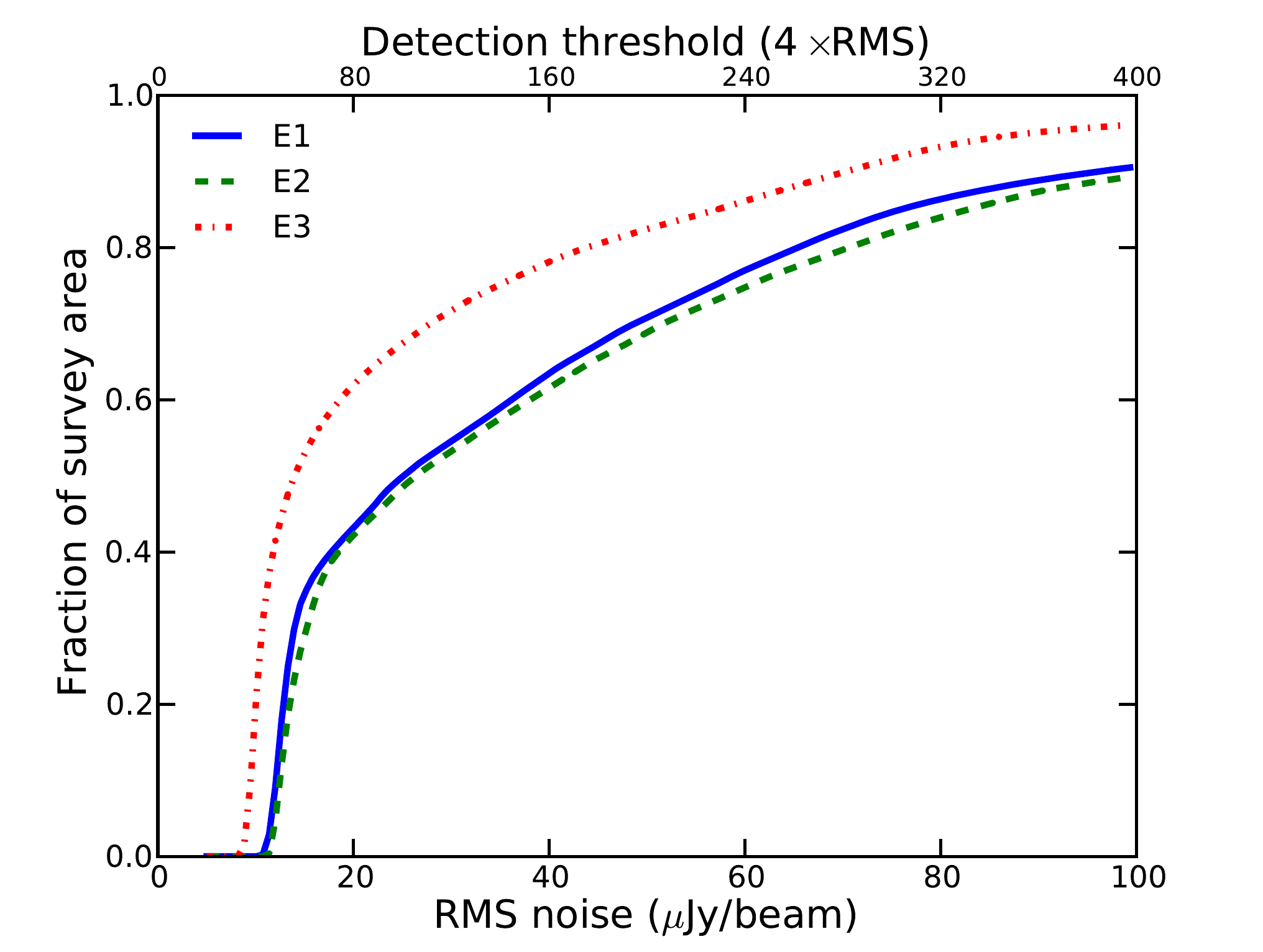}
\caption{Cumulative RMS noise across the 0.38 deg$^2$
survey region for the three epochs of observations (E1, E2 and E3). The source detection threshold (4$\sigma$) is shown on the upper x-axis. 
The 50\% completeness over the 0.28 deg$^2$ HAWC region corresponds to an RMS noise of about 16 $\mu$Jy for epochs E1 and E2 and about 10 $\mu$Jy for epoch E3.}
\label{fig:rms}
\end{figure}

Linear mosaicking of the 37 single-pointing images for each epoch was then carried out using {\tt FLATN} in AIPS. 
The mosaics for the first two epochs are 1500x1500 pix$^{2}$ and the third epoch is 3000x3000 pix$^{2}$. 
The primary beam parameters were acquired from EVLA Memo 195\footnote{Perley et al. 2016, \url{https://library.nrao.edu/public/memos/evla/EVLAM_195.pdf}} during the linear mosaicking step. 
Figure~\ref{fig:rms} shows the cumulative RMS noise plots for the three image mosaics.

We followed an identical procedure for the variable source observations. After running the VLA automated calibration pipeline, we split that data set into individual measurement sets. Then we checked the phase calibrator and the flux calibrators for any RFI from antennae or spectral windows, before proceeding with flagging and clipping. For our imaging, we used almost identical parameters, with the change being image size and cell size to accommodate the array configuration change. 

\subsection{Source cataloging, Point Source Selection and Flux density correction}\label{sec:obs_proc.src}
We used the Search and Destroy (SAD) task within AIPS to generate 4$\sigma$ catalogs for each of our three image mosaics. 
These catalogs contain around 360 sources for epochs E1 and E2 and around 780 sources\footnote{{The increase in the number of sources detected is due to the reduced image noise in E3 and sources (AGN) from E1/E2 being resolved into doubles in E3 (due to the factor of 2 increase in angular resolution between D and C array configurations).}} for epoch E3. 
For a source beyond 200 Mpc, we do not expect contaminating radio emission from any putative host galaxy \citep{hoto2016,2018ApJ...857..143M} of the candidate merger, and hence we shortlisted only the point sources for the transient and variability search.
Our criteria for selecting point-like sources and rejecting false positives (resulting from image artifacts around bright sources) were the same as those used in some previous works \citep{mooley2016,2019MNRAS.490.4898H}:
\begin{itemize}
\item \texttt{BMAJ}/1.7$<$MAJ$<$1.7$\times$\texttt{BMAJ}
\item \texttt{BMIN}/1.7$<$MIN$<$1.7$\times$\texttt{BMIN}, 
\item \texttt{BMAJ}/\texttt{BMIN}$<$2.5
\item $0.67<{\rm Flux/Peak}<1.5$
\end{itemize}
where \texttt{BMAJ} and \texttt{BMIN} are the major and minor axes of the synthesized beam, MAJ, MIN, Flux and Peak are the major and minor axes of the fitted Gaussian and the integrated and peak flux densities as reported by SAD. 
{The first three criteria are motivated by thorough inspection of archival VLA images and source catalogs, and help in differentiating side lobes (false-positives) and spike-like imaging artifacts seen occasionally around bright sources in VLA images. 
They also allow extended source rejection.
The fourth is a simplified criterion for differentiating between resolved and point-like sources (see Figure 9 of \cite{smolcic2017}).}
We then generated a single point-source catalog (PSC) by merging the list of point-like sources for all the three epochs.
The PSC had tens of sources that were present in epoch E3 at the 4--5$\sigma$ level and absent in epochs E2 and E3.
We rejected these sources as false positives after inspecting both the catalog and image mosaic as being due to noise/imaging artifacts, and compiled the final PSC containing 165 sources. 
For all sources in the PSC we plotted a histogram of the ratio of peak flux densities between E1--E2 and E1--E3, and found that flux multiplicative factors of 0.94 and 1.1 were necessary for epochs E2 and E3 respectively in order to make the histograms centered on unity.
We therefore corrected all peak flux densities in E2 and E3 accordingly in the PSC.
\begin{figure}
\centering
\includegraphics[width=3.5in]{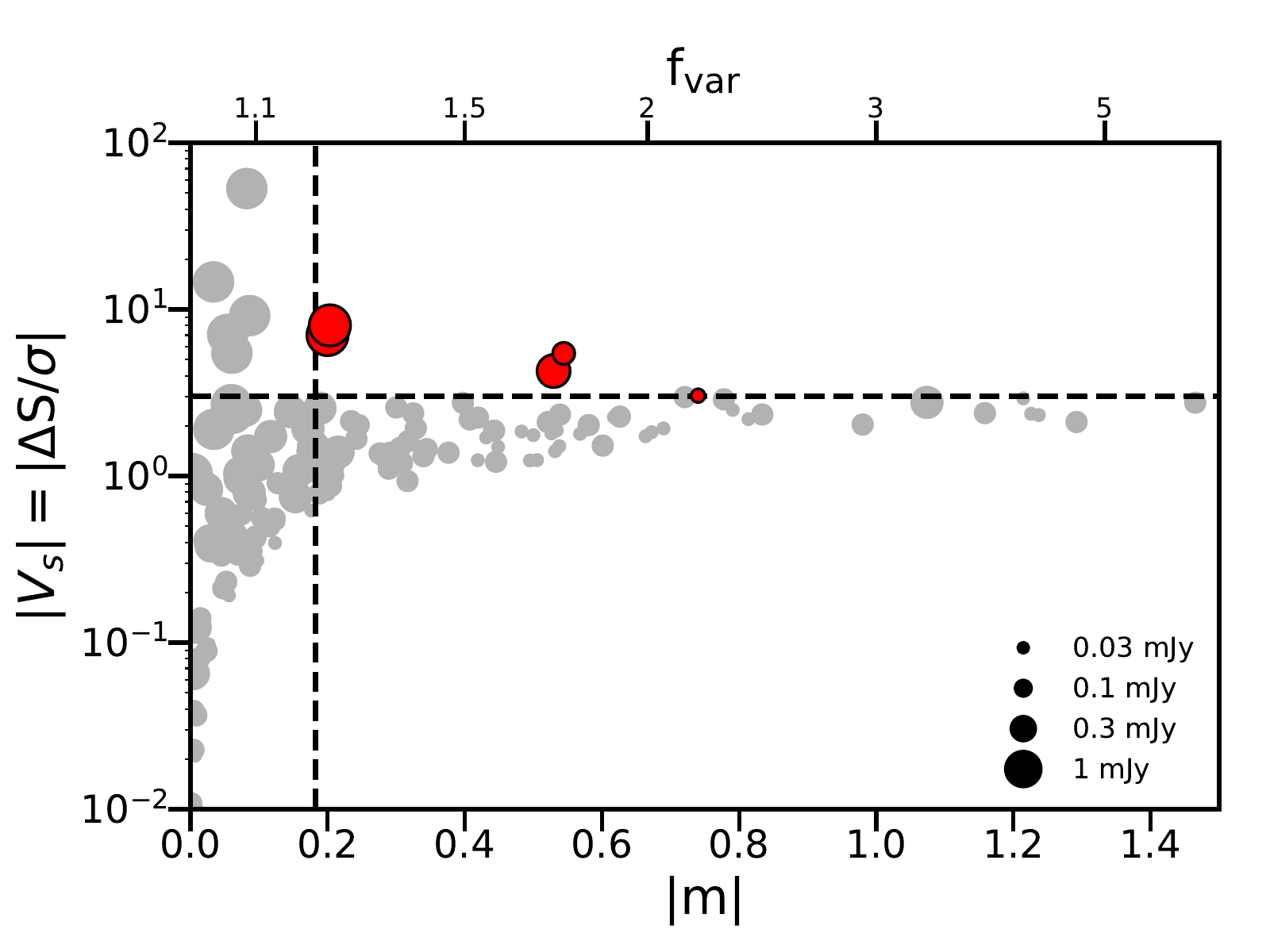}
\caption{Variability statistic ($V_s$) versus the modulation index ($m$) for the 165 sources in our point-source catalog (PSC). 
Grey points indicate sources (from the E1--E2 and E1--E3 comparisons) that are not significant variables.
The red points are the selected variables between E1--E3 (5 sources, see Table~\ref{tab:vartransum}).
No significant variables were found in the E1--E2 comparison.
The black dashed lines indicates the variability selection criteria in $|V_s|$ and $|m|$ (see \S\ref{sec:var_tran}).
The flux densities of the sources defines the marker size (shown in the legend).
Top horizontal scale is the fractional variability $f_{\rm var}$, defined as the ratio of the flux densities between two epochs being compared.
See \S\ref{sec:var_tran} for details.}
\label{fig:var}
\end{figure}
\begin{table*}
\scriptsize
\centering
\caption{Summary of variables sources.}
\label{tab:vartransum}
\begin{tabular}{lllllllllllll}
\hline\hline 
Name & RA    & DEC   & S1    & S2    & S3 & S$_F$ & m & Vs & $\alpha_{4.9}^{7.0}$(E3)  & Host Ident. & r & phot-z\\
(JAGWAR J...) & (deg) & (deg) & ($\mu$Jy) & ($\mu$Jy) &	($\mu$Jy) & ($\mu$Jy) &	& & & & (mag)   \\
\hline
\multicolumn{13}{c}{E1--E2 comparison; Timescale $<$1 week}\\
\hline
\multicolumn{13}{c}{None}\\
\hline
\multicolumn{13}{c}{E1--E3 comparison; Timescale $<$4 months}\\
\hline
J213250+051359& 323.20873 & 5.23317 &  263$\pm$21 & 255$\pm$23 & 139$\pm$15 & 134$\pm$15 & 0.53 & 4.3& $-$1.0$\pm$0.4 & LIRG & 17.5 & 0.09$\pm$0.02\\
J213317+052104 & 323.32286 & 5.35130 &  742$\pm$15 & 881$\pm$16 & 552$\pm$12 & 558$\pm$14 & 0.20 & 7.0 & $-$0.5$\pm$0.1 & Spiral/Ellip. & 22.4 & 0.97$\pm$0.11\\
J213341+051946 & 323.42447 & 5.32948 & 44$\pm$13 &  49$\pm$12 &  87$\pm$11 & 58$\pm$9 & $-$0.74 & $-$3.0 & $-$1.2$\pm$0.8 & LIRG/Spiral & $>$22.7 & \ldots\\ 
J213407+051800 & 323.53327 & 5.30004 &  808$\pm$15 & 956$\pm$15 & 599$\pm$11 & 830$\pm$15 & 0.20 & 8.0 & $-$0.4$\pm$0.1 & \ldots & $>$22.7 & \ldots\\ 
J213453+052633 & 323.72386 & 5.44276 &  124$\pm$13 & 123$\pm$13 & 197$\pm$11 &244$\pm$15 & $-$0.54 & $-$5.4 & +0.3$\pm$0.3& LIRG & 22.1 & 0.61$\pm$0.09 \\ 
\hline
\multicolumn{13}{p{7in}}{Notes: By investigating using two-epoch variability (with Epoch 1 as our reference epoch), we found no variable sources over E1-E2 and 5 variable sources over E1-E3. The flux of each JAGWAR source is reported in the columns S1, S2, S3, and S$_F$, corresponding to each epoch and the follow up observation. The modulation index ($m=\Delta{\rm S}/\bar{\rm S})$ and the Variability Index ($V_s=\Delta S/\sigma$) for each source are calculated and provided in columns m and $V_s$. The host identities were determined using the WISE colors \citep{wright2010} calculated from the ALLWise Catalog catalog \citep{2013yCat.2328....0C} after checking whether the source matched any AGN sources. The r-band magnitude and the spectroscopic redshift were acquired from SDSS DR14. J213341+051946 was initially not detected in Epoch 1 so no in-band spectral index was calculated for Epoch 1.}\\
\end{tabular}
\end{table*}


\section{Transient and Variability Search}\label{sec:var_tran}

We used the PSC from \S\ref{sec:obs_proc.cal} to carry out a search for transients sources which appeared or disappeared in one or more of the three epochs. 
No transients were found to a 4$\sigma$ limit of $\sim$75 $\mu$Jy (mean completeness threshold for the merged catalog over 3 epochs and 100 deg$^2$).

Following \cite{mooley2016} we used PSC to also investigate two-epoch variability using the the variability statistic, $V_s=\Delta S/\sigma$ and modulation index $m=\Delta{\rm S}/\bar{\rm S}$, where S is the flux density, 
$\bar{\rm S}$ is the average flux density over the two epochs being compared, $\Delta$S is the flux density difference, and $\sigma$ is the RMS noise.
We used epoch E1 as the reference and performed the following two-epoch comparisons: E1--E2 and E1--E3.  
Significant variables were identified as those sources having $|V_s|$ larger than three (corresponding to Gaussian equivalent of approximately 3$\sigma$, i.e. a chance probability of finding 3 variables out of 1000 sources; this ensures that less than one false positive will be detected as a variable source in our search, assuming Gaussian statistics) and the absolute value of the modulation index, $|m|$, larger than 0.18 (i.e. a fractional variability, $f_{\rm var}>1.2$; this was chosen bearing in mind that flux correction factors of up to 10\% were applied to the flux densities within the PSC and that our flux scale is accurate to only $\sim$5\%). 

The plot of the variability statistic versus the modulation index is shown in Figure~\ref{fig:var}. We found no significant variable sources in the E1--E2 comparison (probing a timescale of $<$1 week) and 5 significant variables in the E1--E3 comparison (probing a timescale of $<$4 months). This indicates that $<$2\% of the persistent sources are variable on $<$1 week timescale and $3.0\pm1.3$\% of the persistent sources are variable on timescales of week-months. This level of variability is typical for the radio sky \citep[e.g.][]{2003ApJ...590..192C,bell2015,mooley2016,2016MNRAS.461.3314H}, and at these frequencies it is attributed to normal activity from active galactic nuclei \citep[AGN;][]{2019MNRAS.490.4024R}.

\begin{figure}
\centering
\includegraphics[width=4in,angle=0]{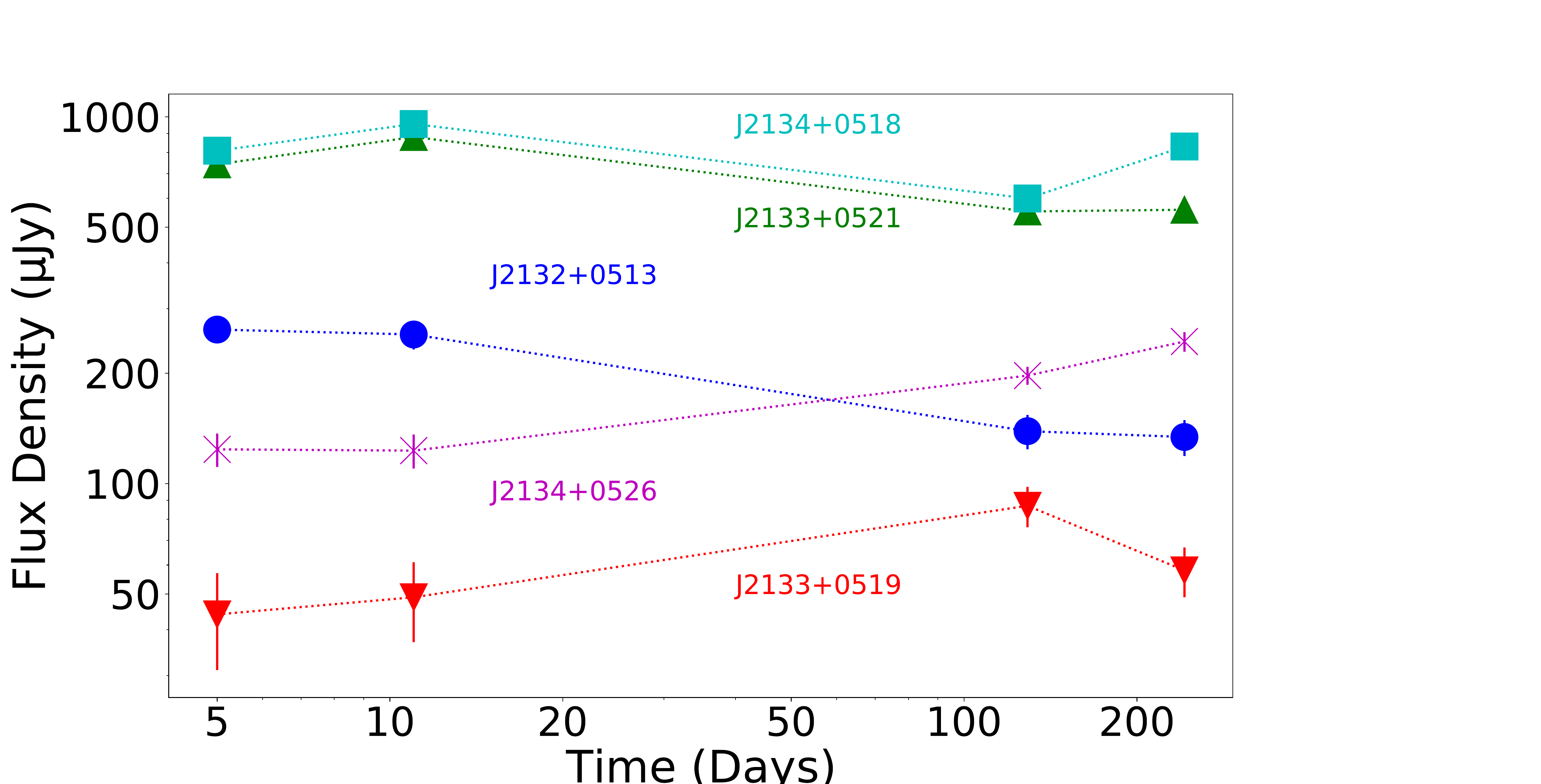}
\caption{Radio light curves at 6 GHz for the five variable sources identified in Table \ref{tab:vartransum}. The x-axis gives the time since merger. Flux densities for each source are shown different shapes (circle, triangle, inverted triangle, square and star).}
\label{fig:LC}
\end{figure}

Next we examined the properties of these five variable sources for any indication that they may be long-lived transients related to S191216ap and not just background AGN. Their properties are summarized in Table~\ref{tab:vartransum}. Radio data include the sources positions (RA/DEC), the flux densities for all three epochs (S1, S2, S3), the modulation index (m), the variability statistic ($V_s$), and the in-band spectral indices ($\alpha$) for the first (E1) and third epochs (E3). Also included are the results of WISE counterpart source matching \citep{wright2010} in which we attempted to classify the radio source using WISE colors from the AllWise Catalog \citep{2013yCat.2328....0C} and \citet{2014MNRAS.442.3361N}. Finally, we used the latest release of the Sloan Digital Sky Survey (SDSS DR14) in which we record the r-band magnitude and the photometric redshifts, where available, for each WISE source.

We found WISE counterparts for four JAGWAR sources, J213250+051359, J213317+052104, J213341+051946 and J213453+052633. The remaining JAGWAR source J213407+051800, is located 11.6 arcsec from AllWISE J213407.27+051804.8, with rmag=20.8 and a photometric redshift of 0.44. None of these five sources are in the WISE AGN catalog of \citet{2018ApJS..234...23A}. From their WISE colors we deduce from \citet{wright2010} that the putative hosts are variously luminous infrared galaxies (LIRGs), spirals and/or elliptical galaxies (see Table~\ref{tab:vartransum}). Two sources, J213317+052104 and J213453+052633, have photometric redshift values that place the host galaxy far beyond the luminosity distance of S191216ap.

The light curves are shown in Figure \ref{fig:LC}. They include the VLA follow-up observations from day 242 (see Table~\ref{tab:observations}). We detected all of five sources at integrated flux densities of 134$\pm$15\,$\mu$Jy, 558$\pm$14\,$\mu$Jy, 58$\pm$9\,$\mu$Jy, 830$\pm$15\,$\mu$Jy, 244$\pm$10\,$\mu$Jy, respectively (in the order listed in Table~\ref{tab:vartransum}). With only four epochs, it is not easy to make any definitive statements, but they are consistent with fluctuations from persistent radio sources and none of the sources show the sharp rise and decay pattern of an afterglow. J213453+052633 exhibits a rise, nearly doubling during the 242 days. However, the photometric redshift of its host galaxy rules out an association with S121916ap. The brightest radio source J213407+051800 is also detected in the first-look images of the VLA Sky Survey (VLASS) with a 3 GHz flux density of 569$\pm$122\,$\mu$Jy, from data taken 20 October 2017 \citep{2020PASP..132c5001L}. None of the other radio sources were detected in the VLASS with 3$\sigma$ limits $\leq$375\,$\mu$Jy. Finally we note that in the higher resolution follow-up observations, JAGWAR J213250+051359 clearly shows a core-jet morphology.

Based on the host galaxy classifications and redshifts, plus the amplitude, timescale and fractional variation level, persistence of the radio emission and radio source morphology, it seems likely that these variable radio sources are background low-luminosity AGN and likely unrelated to S191216ap.

\section{Discussion and Future Prospects}\label{sec:discussion}

Stellar mass binary black hole mergers were not widely expected to generate electromagnetic (EM) counterparts, so early predictions focused on EM signatures from binary NSs and BH-NS binaries \citep[e.g.][]{metzger2012,2013ApJ...767..124N}. Nevertheless, observations were still undertaken to search for incoherent radio emission in the first two science runs of LIGO \citep{2016ApJ...829L..28P, 2018ApJ...857..143M,2019ApJ...884...16A}. Another promising avenue has been the search for prompt {\it coherent} radio emission on timescales of minutes to hours post-merger, using wide-field low frequency arrays \citep{2015ApJ...812..168Y, 2016PASA...33...50K, 2019ApJ...877L..39C, 2019MNRAS.489L..75J,2019MNRAS.489.3316R}. Renewed impetus for afterglow searches came following the nominal detection (2.9$\sigma$) of a gamma-ray flare from {\it Fermi-GBM} detection toward GW\,150914 \citep{2016ApJ...826L...6C,2018ApJ...853L...9C}. This stimulated a number of theoretical investigations for stellar mass BBHs \citep[e.g.,][]{Loeb2016,Perna2016,Woosley2016,Zhang2016}, which in several cases predict specific EM signatures that are testable with follow-up radio observations \citep{2016PTEP.2016e1E01Y,2016ApJ...825L..24M,Perna2019}.

As noted earlier (\S\ref{sec:intro}), S191216ap has been classified as a BBH merger event seen by LIGO-Virgo, which had a coincident neutrino detection from IceCube, and a possible detection of a EM counterpart at TeV energies from HAWC. We have carried out a search for a radio transient within the HAWC error circle for three epochs covering timescales between 4 days and 4 months, at a wavelength of 6 cm (\S\ref{sec:obs_proc.cal}). While no radio transients were discovered, our flux density upper limits represent a considerable improvement over past BBH radio afterglow searches. Our 4$\sigma$ limits\footnote{{This is the mean sensitivity across the HAWC region.}} of 75 $\mu$Jy for a radio afterglow from S191216ap is an improvement over the 4$\sigma$ radio limits of 600 $\mu$Jy limits for GW151226 \citep{2018ApJ...857..143M} and 180 $\mu$Jy for GW170608 \citep{2019ApJ...884...16A}, two BBH events that occurred during the O2 and O3 Virgo science runs, respectively. Factoring in the luminosity distance of S191216ap, the radio luminosity is 1.2$\times{10}^{28}$ erg s$^{-1}$ Hz$^{-1}$, or approximately 5--10 times deeper than these previous BBH radio afterglow searches. 

\begin{figure}
\centering
\includegraphics[height=9cm,angle=-90]{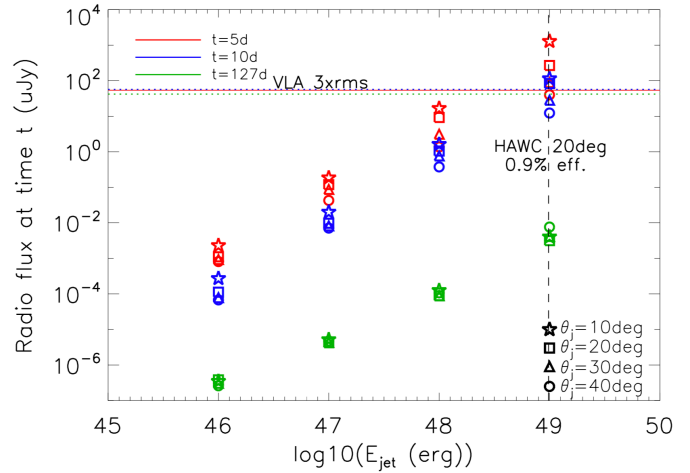} 
\caption{Sensitivity reached in our VLA follow-up (horizontal lines; see also see Table \ref{tab:observations}) with models  for potential BBH radio afterglows as seen from an on-axis observer \citep[symbols;][]{Perna2019}. See Section \ref{sec:discussion} for details. }
\label{fig:BBH}
\end{figure}

Given the absence of a radio transient for this GW/neutrino/TeV candidate event, we next look at how our radio limits can be used to set limits on any putative afterglow from S191216ap. We adopt the model of \citet{Perna2019}, in which a jet is formed during the BBH merger that propagates freely, without any baryonic contamination from tidal disrupted material, until it interacts with the surrounding interstellar medium and generates afterglow emission. In Figure \ref{fig:BBH} we compare the sensitivity reached in our VLA follow-up (horizontal lines; see also see Table \ref{tab:observations}) with models for potential BBH radio afterglows as seen from an on-axis observer. The model predictions at timescales comparable to those of our radio follow-up are shown for different values of the total kinetic energy in the jet ($E_{\rm jet}$) and of the jet half-opening angle $\theta_j$ \citep[symbols;][]{Perna2019}. We have rescaled the model values provided at 1.4\,GHz by \citet{Perna2019} to 6\,GHz assuming an optically thin spectral index of 0.65 (which is consistent with the model itself between radio and optical frequencies). These model predictions also assume $\Gamma=100$, $\epsilon_e=0.03$, $\epsilon_B=0.01$, and $n_{\rm ISM}=0.01$\,cm$^{-3}$ \citep{Perna2019}. We note that the observed fluxes scale as $n^{5/14}$ for a fully radiative blast wave and as $n^{1/2}$ for an adiabatic evolution. Thus, generally speaking, higher densities imply larger fluxes \citep{Perna2019}. Given the tentative HAWC sub-threshold event at 1 TeV, we focus our comparison on the $\Gamma=100$ (highly relativistic jet) models presented in \citet{Perna2019} rather than on the $\Gamma=10$ (mildly relativistic jet) case.

As evident from this Figure \ref{fig:BBH}, our radio follow-up campaign was sensitive to only the most optimistic EM counterpart models in terms of energy coupled to a relativistic ejecta.  For reference, in Figure \ref{fig:BBH} we also mark with a vertical dashed line an order-of-magnitude energy estimate derived from the flux density measurement of the HAWC sub-threshold event at 1\,TeV as \citep{2019GCN.26472....1H}:
\begin{eqnarray}
    \nonumber E_{\rm HAWC}\sim (1\,{\rm TeV})^2 \times  7.3\times10^{-9}\,{\rm TeV^{-1}cm^{-2}\,s^{-1}}\\\times10\,{\rm s}\times \frac{4\pi d^2_L(1-\cos(20\,{\rm deg}))}{\xi},
\end{eqnarray}
where we neglect redshift corrections, we set $d_L=320$\,Mpc, have assumed a high-energy signal duration of 10\,s, a jet opening angle of 20\,deg, and an efficiency of $\xi=0.9$\% for the conversion of ejecta kinetic energy into prompt emission energy at 1\,TeV. This value of the efficiency is chosen so that the flux density measured by HAWC is consistent with the hypothesis of a BBH jet with kinetic energy of $10^{49}$\,erg (and opening angle of 20\,deg), which is the minimum energy value for which our radio upper-limits are constraining of the model predictions. 

Our work has a number of limitations. In particular, our VLA imaging campaign focused on the area defined by the HAWC error circle \citep{2019GCN.26472....1H}. If the gamma-ray emission (or neutrino detection) were were not deemed significant, then the multi-messenger aspect of this event would be in question, and we would have surveyed less than 0.1\% of the full error region of S191216ap. Furthermore, even though the observations presented here are an improvement in sensitivity over earlier LIGO-Virgo science runs, the current radio limits are still not sufficiently constraining on {\it all} predicted EM signatures from BBHs. For example, radio methods are also not particular powerful for finding EM signatures from BBH mergers in accretion disk environments of supermassive BHs \citep{Stone2017,Bartos2017,2019ApJ...884L..50M,Tagawa2020}, a model invoked to argue that a peculiar optical flaring AGN was the counterpart of the GW event S190521g \citep{2020PhRvL.124y1102G}. Radio is well-suited for identifying transients, but radio variability is most commonly ascribed to regular AGN variability. Nonetheless, the phase space remains large for both prompt and longer-term searches for coherent and incoherent radio emission, respectively. Such searches should continue during the fourth GW science run which should include the Kamioka Gravitational Wave Detector (KAGRA) detector, with array improvements in both sensitivity and localization \citep{2018LRR....21....3A}. The detection of radio afterglow from a BBH would revolutionize the studies of such objects in much the same way that the multi-messenger studies of GW170817 advanced our understanding of binary neutron stars.


\acknowledgments{\it The National Radio Astronomy Observatory is a facility of the National Science Foundation operated under cooperative agreement 
by Associated Universities, Inc.
We would like to thank the NRAO staff, especially Amy Mioduszewski, Heidi Medlin, Drew Medlin, Tony Perreault and Abi Smoake for help with observation scheduling and computing. 
K.P.M. is currently a Jansky Fellow of the National Radio Astronomy Observatory.
K.P.M and G.H. acknowledge support from the National Science Foundation Grant AST-1911199. A.C., A.B., and D.B. acknowledge support from the National Science Foundation via the CAREER grant \#1455090.
DK is supported by NSF grant AST-1816492.

}


\end{document}